\documentclass[sort&compress,12pt]{article}%
\usepackage{amssymb}
\usepackage{amsfonts}
\usepackage{amsmath}
\usepackage{breqn}
\usepackage{graphicx}
\usepackage[T1]{fontenc}
\usepackage[utf8]{inputenc}
\begin{document}

\title{Exact Noncommutative Two-Dimensional Hydrogen Atom}
\author{B.C. Wang$^{a}$, E.C. Brenag$^{b}$ R.G.G. Amorim$^{a,c,d}$, V.C. Rispoli$^{a}$, S.C. Ulhoa$^{c,d}$\\${}^{a}$ Universidade de Bras\'{\i}lia, Faculdade Gama,\\72444-240, Bras\'{\i}lia, DF, Brazil\\${}^{b}$ Universidade de Bras\'{\i}lia, Faculdade de Tecnologia,\\70910-900, Bras\'{\i}lia, DF, Brazil\\${}^{c}$International Centre for Condensed Matter Physics$,$\\Instituto de F\'{\i}sica, Universidade de Bras\'{\i}lia,\\70910-900, Bras\'{\i}lia, DF, Brazil\\${}^{d}$Canadian Quantum Research Center,\\
204-3002 32 Ave Vernon, BC V1T 2L7  Canada\\}
\maketitle

\begin{abstract}
In this work, we present an exact analysis of two-dimensional noncommutative hydrogen atom. In this study, it is used the Levi-Civita transformation to perform the solution of the noncommutative Schrödinger equation for Coulomb potential. As an important result, we determine the energy levels for the considered system. Using the result obtained and experimental data, a bound on the noncommutativity parameter was obtained. 


\end{abstract}

\section{Introduction}

The concept of noncommutativity in physical theories was formally introduced by Snyder in 1947 \cite{jackiw, snyder1, snyder2}. In a seminal paper, Snyder stated that spatial coordinates would not commute with each other at small distances. In this sense, a new paradigm was proposed in which the space-time should be understood as a collection of tiny cells of minimum size, where there is no such idea of a point. So far, once the minimum size is reached, in the realm of some high energy phenomenon, the position should be given by the noncommutative coordinate operators. As a direct consequence, it would be impossible to precisely measure a particle position. Over the last
years, the interest of the scientific community on noncommutative
geometry has increased due to works on non-abelian
theories~\cite{chans}, gravitation~\cite{kalau, kastler, connes1},
standard model~\cite{connes2, varilly1, varilly2} and quantum
Hall effect~\cite{belissard}. More recently, the discovery that the
dynamics of an open string can be associated with noncommutative
spaces has contributed to the last revival of noncommutative
theories~\cite{witten}.

From the mathematical point of view, the simplest algebra obeyed by
the coordinates operators $\widehat{x}^{\mu}$ is
$$
[\widehat{x}^{\mu},\widehat{ x}^{\nu}]=i\Theta^{\mu\nu}\,,
$$
where $\Theta^{\mu\nu}$ is a skew-symmetric constant tensor called noncomutativity parameter. It is worth to point out that the mean values of the position operators
do correspond to the actual position observed, thus it is said that
such operators are Hermitian ones. It is well known in quantum
mechanics that a noncommutative relation between two operators lead
to a specific uncertain relation, hence the above expression yields
$$
\Delta \widehat{x}^{\mu}\Delta \widehat{x}^{\nu}\geq
\frac{1}{2}|\Theta^{\mu\nu}|\,,
$$
in which implies a set of uncertainty relations between position coordinates analogous to the Heisenberg uncertainty principle.
Following the ideas introduced by Snyder, it is possible to associate
the minimum size with a distance of the $\sqrt{|\Theta^{\mu\nu}|}$
order of magnitude. Thus the noncommutative effects turn out to be
relevant at such scales. Usually the noncommutativity is introduced
by means of the Moyal product defined as \cite{snyder1}

	\begin{equation}
		f(x) \star g(x) \equiv \exp
		\left(  {i \over 2} \Theta^{\mu\nu}  {\partial \over \partial x^\mu}
		{\partial \over \partial y^\nu} \right) f(x) g(y) |_{y \rightarrow x}, 
	\end{equation}
with a constant $\Theta^{\mu\nu}$. Then, the usual product is replaced by the Moyal product in the classical Lagrangian density. In similar perspective, the noncommutative quantum mechanics is introduced by imposing further commutation relations between the position coordinates themselves.

In this perspective, we aim to apply the ideas about noncommutativity of space in the hydrogen atom. The hydrogen atom is an electrically neutral atom with  a positively charged proton and a negatively charged electron bounded to the nucleus. This system plays a significant role in quantum mechanics and field theory. There are many good reasons to address the hydrogen atom \cite{hyd1}. As an example, the hydrogen atom with high precision measurements in atomic  transitions is one of the best laboratories to test quantum electrodynamics theory \cite{hyd2}. Other applications of the hydrogen atom appear in many occasions, such as to examine the constancy of fine structure constant over a cosmological time scale \cite{hyd3}. In this paper, we analyze the two-dimensional noncommutative hydrogen atom. A two-dimensional hydrogen atom can be defined as a system in which the motion  of the electron around the proton is constrained to be planar. Then, in this work, we consider that this plane is noncommutative. As a practical example, a semiconductor quantum-well under illumination is a quasi-two-dimensional system, in which photoexcited  electrons and holes are essentially confined to a plane \cite{hyd4,hyd5}. There are many works that consider the hydrogen atom in a noncommutative context, but they present disagreement in results. Our method presents an approach using the Levi-Civita mapping, which allows an exact treatment.

This paper is organized as follows. In section 2, we present the mathematical framework of the two-dimensional hydrogen atom. The Levi-Civita mapping is presented in section 3. In section 4, we obtain the solution and spectrum for the noncommutative hydrogen atom.  Finally,
in the section 5, we present our concluding remarks.
\section{Mathematical Framework of Noncommutative Two-Dimensional Hydrogen Atom}
The Hamiltonian that define two-dimensional hydrogen atom is given by
\begin{equation}\label{h1}
H=\frac{1}{2m}(p_{x}^2+p_{y}^2)-\frac{k}{(x^2+y^2)^{1/2}},
\end{equation}
where $p_{x}$ and $p_y$ stands for the momentum of the electron in directions $x$ and $y$, respectively; $x$ and $y$ represents the electron coordinates and the constant $k=\frac{e^{2}}{4\pi\epsilon_o}$, where $e$ is the elemental electrical charge and $\epsilon_o$ is the vacuum electrical permissivity. To quantize the Hamiltonian given in Eq.(\ref{h1}), as usual, the momentum operators are given by $\widehat{p}_{x}=-i\hbar\frac{\partial}{\partial x}$ and $\widehat{p}_{y}=-i\hbar\frac{\partial}{\partial y}$, where $\hbar=h/2\pi$ and $h$ is the Planck constant.

In the noncommutative perspective, we define the following positions operators
\begin{equation}\label{h2}
\widehat{x}=x+\frac{i\Theta}{2}\frac{\partial}{\partial y},
\end{equation}
\begin{equation}\label{h2}
\widehat{y}=y-\frac{i\Theta}{2}\frac{\partial}{\partial x},
\end{equation}
in which $\Theta$ is the noncommutativity parameter in Cartesian coordinates. We notice that
\begin{equation}\label{h3}
[\widehat{x},\widehat{y}]=i\Theta,
\end{equation}
as expected.

However, the treatment of the Hamiltonian given in Eq.(\ref{h1}) is difficult because of the operators in the denominator of the potential energy term. For this reason, in the next section we present a transformation that put the system in a more suitable way. 

\section{Levi-Civita Mapping}

The Levi-Civita (also known as Bohlin) transformation is a parabolic coordinates mapping that is capable to convert the planar Coulomb problem in a two-dimensional harmonic oscillator \cite{levicivita1, levicivita2,campos,kliber}. It is a $\mathbb{R}^2\rightarrow\mathbb{R}^2$ surjection defined by
\begin{eqnarray}
x&=&u^2-v^2,\nonumber\\
y&=&2uv. \label{h4}
\end{eqnarray}
Given Eqs.(\ref{h4}) it is immediate to conclude that
\begin{eqnarray}
\frac{\partial}{\partial u}=2\left(u\frac{\partial}{\partial x}+ v\frac{\partial}{\partial y}\right),\nonumber\\
\frac{\partial}{\partial v}=2\left(-v\frac{\partial}{\partial x}+ u\frac{\partial}{\partial y}\right), \label{h4n1}
\end{eqnarray}
and, by inversion, 
\begin{eqnarray}
\frac{\partial}{\partial x}=\frac{1}{2(u^2+v^2)}\left(u\frac{\partial}{\partial u}-v\frac{\partial}{\partial v}\right),\nonumber\\
\frac{\partial}{\partial y}=\frac{1}{2(u^2+v^2)}\left(v\frac{\partial}{\partial u}+ u\frac{\partial}{\partial v}\right). \label{h4n2}
\end{eqnarray}
As a direct consequence of Eq.(\ref{h4n2}), the momentum operators can be rewritten in this new coordinate system as
\begin{eqnarray}
p_x&=&\frac{p_uu-p_vv}{2(u^2+v^2)}\nonumber\\
p_y&=&\frac{p_uv+p_vu}{2(u^2+v^2)}.\label{h4n3}
\end{eqnarray}
It should be noted that the Levi-Civita mapping is a canonical transformation \cite{campos}.

Applying Eqs.(\ref{h4}) and Eqs.(\ref{h4n3}) in Eq.(\ref{h1}), we obtain the following transformed Hamiltonian 
\begin{equation}\label{h5}
H=\frac{1}{2m}\left[\frac{p_u^2+p_v^2}{4(u^2+v^2)}\right]-\frac{k}{(u^2+v^2)}.
\end{equation}
Finally, the hypersurface defined by $H=E$ is given by
\begin{equation}\label{h6}
\frac{1}{2m}(p_u^2+p_v^2)-4E(u^2+v^2)=4k.
\end{equation}
Equation (\ref{h6}) is the main result of this section and is the one to be used from now on. 
\section{Analysis of Two-Dimensional Hydrogen Atom}
Applying  the following set of operators in Eq.(\ref{h6})
\begin{eqnarray}\nonumber
\widehat{u}&=&u+\frac{i\theta}{2}\frac{\partial}{\partial v},\nonumber\\
\widehat{v}&=&v-\frac{i\theta}{2}\frac{\partial}{\partial u},\nonumber\\
\widehat{p}_{u}&=&-i\hbar\frac{\partial}{\partial u},\nonumber\\
\widehat{p}_{v}&=&-i\hbar\frac{\partial}{\partial v},\nonumber
\end{eqnarray}
we obtain the modified Schrödinger equation as
\begin{equation}\label{h7}
-\left(\frac{\hbar^2}{2m}-E\theta^2\right)\left(\frac{\partial^2\psi}{\partial u^2}+\frac{\partial^2\psi}{\partial v^2}\right)-4E\left[(u^2+v^2)\psi-i\theta\left(v\frac{\partial\psi}{\partial u}-u\frac{\partial\psi}{\partial v}\right)\right]=4k\psi,
\end{equation}
where $\psi\equiv \psi(u,v)$,  $\theta$ is the noncommutativity parameter in parabolic coordinates, $[\widehat{u},\widehat{v}]=i\theta$, $[\widehat{u},\widehat{p}_u]=i\hbar$ and $[\widehat{v},\widehat{p}_v]=i\hbar$. Here is crucial to note that the order of $\theta$ is $\theta \sim \sqrt{\Theta}$, due to Eqs.(\ref{h4}).

The solution of Eq.(\ref{h7}) can be obtained from the following change of variables $z=u^2+v^2$ and then Eq.(\ref{h7}) can be rewritten as
\begin{equation}\label{h8}
z\frac{d^2\psi(z)}{dz^2}+\frac{d\psi(z)}{dz}+\frac{1}{\beta}(Ez+k)\psi(z)=0,
\end{equation}
with $\beta=\left(\frac{\hbar^2}{2m}-E\theta^2\right)$. Defining $\psi(z)=e^{-\lambda z}\phi(z)$, where $\lambda=\sqrt{\frac{-E}{\beta}}$, Eq.(\ref{h8}) can be written as
\begin{equation}\label{h9}
z\frac{d^2\phi(z)}{dz^2}+(1-2\lambda z)\frac{d\phi(z)}{dz}+\left(\frac{k}{\beta}-\lambda\right)\phi(z)=0.
\end{equation}
Performing the change of variable $w=2\lambda z$, we finally obtain
\begin{equation}\label{h10}
w\frac{d^2\phi(w)}{dw^2}+(1-w)\frac{d\phi(w)}{dw}+\frac{1}{2\lambda}\left(\frac{k}{\beta}-\lambda\right)\phi(w)=0.
\end{equation}
Note that Eq.(\ref{h10}) has the following form
\begin{equation}\nonumber
w\phi''+(1-w)\phi'+l\phi=0,
\end{equation}
which is the Laguerre differential equation. If $l$ is an integer $l=0,1,2,3,\ldots$ the solution of Laguerre’s equation is given by Laguerre polynomials $L_l(x)$.
In this sense, we obtain the solution
\begin{equation}\label{h11}
\psi(u,v)=e^{-\lambda(u^2+v^2)}L_l(2\lambda(u^2+v^2)).
\end{equation}
The energy levels can be determined from
\begin{equation}\label{h12}
l=\frac{1}{2}\left(\frac{k}{\lambda\beta}-1\right),
\end{equation}
with $l=0,1,2,3,\ldots$. Using the condition given in Eq.(\ref{h12}), the spectrum can be calculated as
\begin{equation}\label{h12a}
E^2-\frac{\hbar^2}{2m\theta^2}E-\frac{k^2}{n^2\theta^2}=0,
\end{equation}
where $n=2l+1$. Solving Eq.(\ref{h12a}) we obtain the spectrum of noncommutative two-dimensional hydrogen atom
\begin{equation}\label{h13a}
E_n=\frac{\hbar^2}{4m\theta^2}-\frac{\hbar^2}{4m\theta^2}\sqrt{1+\frac{16k^2m^2\theta^2}{\hbar^4n^2}}.
\end{equation}
Considering $\frac{16k^2m^2\theta^2}{\hbar^4n^2} \ll 1$ and using the binomial series $(1+x)^j = 1+jx+j(j-1)x^2/2+\ldots$ we calculate the following approximation 
\begin{equation}\label{h13}
E_n\approx\frac{-me^4}{8\pi^2\epsilon_o^{2}\hbar^2n^2}+\frac{e^8m^3\theta^2}{32\pi^4\epsilon_{o}^4\hbar^6n^4}.
\end{equation}
 Notice that in the limit $\theta \rightarrow 0$, we obtain the same result of the usual two-dimensional hydrogen atom given in the literature \cite{hyd2,hyd3}. Notice also that the first order term in $\theta$ do not contribute to energy of this system.

Then, the noncommutative correction, $
\Delta E_{NC}$, for the energy is given by
\begin{equation}\label{h15}
\Delta E_{NC}\approx\frac{e^8m^3\theta^2}{32\pi^4\epsilon_{o}^4\hbar^6n^4}.
\end{equation}
The result given in Eq.(\ref{h15}) can be used to estimate the bound on the noncommutativity parameter $\theta$. The experimental value for $1S\rightarrow 3S$ frequency transition in the hydrogen atom is $\nu_{1S\rightarrow 3S}=(2 922 743 278 671.6 \pm 0.9) kHz$ \cite{exp}. The uncertainty in this experimental value $\Delta \nu=0.9kHz$ can fix the lower bound on the parameter $\theta$. In this sense, the theoretical value for the error in transition $1S \rightarrow 3S$, denoted by $\Delta E_{1\rightarrow 3}$, is given by
\begin{equation}\nonumber
\Delta E_{1\rightarrow 3}\approx \frac{m^3e^8\theta^2}{32\pi^4\epsilon_{o}^4\hbar^6}\left[\frac{80}{81}\right],
\end{equation}
Using the fact that in the bidimensional case the energy is four times bigger than tridimensional case, then $\frac{\Delta E_{1\rightarrow 3}}{4}= h\nu_{1S\rightarrow 3S}$, where $h$ is Planck constant. So, we have
\begin{equation}\nonumber
\theta\lesssim\left[\frac{324}{5}\frac{\pi^3\epsilon_{o}^4 \hbar^7\Delta \nu}{m^3e^8}\right]^{1/2}.
\end{equation}
Performing all the calculations, we obtain $\theta \lesssim 2.23\cdot 10^{-18} m$. In this case, the bound of the noncommutativity parameter $\Theta $ is   $\Theta \lesssim 4.972\cdot 10^{-36} m^2$. Using the definition of the length scale factor, $\Gamma_{NC}=\sqrt{|\Theta|}$, that is the length scale where the noncommutative effects of the space will be relevant, we found for the considered case $\Gamma_{NC}\lesssim 2.23\cdot 10^{-18}  m $, which is about one thousand times smaller than the proton radius $r_p\approx 0,833\cdot 10^{-15} m$.

\section{Concluding remarks}
Using the Levi-Civita mapping, we treated the non-trivial problem of the noncommutative hydrogen atom. As results, we obtained the solution of Schrödinger equation for this system and calculate the energy levels. Using the spectrum obtained and the experimental data, we estimated the noncommutativity parameter $\Theta$, which has the order of magnitude about $10^{-36} m^2$, and the noncommutative effects will be relevant to lengths smaller than $10^{-18} m$. These results are in agreement with the literature \cite{exp, exp1}. 
\section*{Acknowledgements}

This work was partially supported by CNPq of Brazil.


\begin{thebibliography}{99}                           

\bibitem{jackiw} R. Jackiw. 
Nucl. Phys. Proc. Suppl. 108 (2002) 30.

\bibitem{snyder1} H. S. Snyder. 
Phys. Rev. 71 (1947) 38.

\bibitem{snyder2} H. S. Snyder. 
Phys. Rev. 72 (1947) 68.

\bibitem{chans} A. H. Chanseddine, G. Felder, J. Frohlich. 
Commun. Math. Phys. 155 (1993) 205.

\bibitem{kalau} W. Kalau, M. Walze. 
J. Geom. Phys. 16 (1955) 327.

\bibitem{kastler} D. Kastler. 
Commun. Math Phys. 166 (1995) 633.

\bibitem{connes1}  A. H. Chanseddine, A. Connes. 
Commun. Math Phys. 186 (1997) 731.

\bibitem{connes2} A. Connes, J. Loot. 
Nucl. Phys. Proc. Suppl. 18B (1991) 29.

\bibitem{varilly1} J. C. Varilly, J. M. Garcia-Bondia. 
J. Geom. Phys. 12 (1993) 223.

\bibitem{varilly2} C. P Martin, J. C. Varilly, J. M. Garcia-Bondia. 
Phys. Rept. 294 (1998) 363.

\bibitem{belissard} J. Belissard, A. van Elst, H. Schulz-Baldes. 
J. Math. Phys. 35 (1994) 53.

\bibitem{witten} N. Seiberg, E. Witten, JHEP \textbf{9909} (1999) 32, hep-th/9908142.

\bibitem{jack1} R. Jackiw, V.P. Nair. Phys. Lett. B \textbf{480}, 237 (2000).

\bibitem{hyd1} D. Palmer, \emph{Hydrogen in the Universe}, NASA (2008).

\bibitem{hyd2} B. Zaslow, M.E. Zandler, Am. J. Phys \textbf{35} (1967) 1118..

\bibitem{hyd3} X.L. Yang, \emph{et. al.}, Phys. Rev. A \textbf{43} (1991) 1186.

\bibitem{hyd4} X.L. Yang, M. Lieber, F.T. Chan. Am. J. Phys. \textbf{59} (1991) 231.

\bibitem{hyd5} M. Chaichian, M.M. Sheik-Jabbari, A. Turianu. Eur. Phys. J. C \textbf{36} (2004) 251.

\bibitem{levicivita1} T. Levi-Civita, Opere Mat.\textbf{2} (1906) 417.

\bibitem{levicivita2} T. Levi-Civita,
 Acta Mathematica \textbf{42} (1920) 99.

\bibitem{campos} P. Campos, M.G.R. Martins, J.D.M. Viana. Phys. Lett. A \textbf{381} (2017) 1129.

\bibitem{kliber} M. Kliber, T. Negadi. Croat. Chem. Acta \textbf{57} (1984) 1509.

\bibitem{exp} H. Fleurbaey, \emph{et. al.} Phys. Rev. Lett. \textbf{120} (2018) 183001.

\bibitem{exp1} T.W. Hansch \emph{et al}, Philos. Trans. R. Soc. A \textbf{363} (2005) 2155.

\bibitem{nc1} A. Stern, Phys. Rev. Lett. \textbf{100} (2008) 061610. 
\end{thebibliography}
\end{document}